\documentclass[review]{elsarticle}
\usepackage{amsmath,amssymb}
\usepackage{lipsum}
\usepackage{bm}
\usepackage{graphicx}
\usepackage{graphics}
\usepackage{amsmath}
\usepackage{array}
\usepackage[caption=false]{subfig}
\usepackage{xcolor}
\usepackage{subfig}
\usepackage{hyperref}
\usepackage[capitalise]{cleveref}
\usepackage{textcomp}

\usepackage{epstopdf}
\usepackage{cases}
\usepackage{amssymb}
\usepackage{multirow}
\usepackage{siunitx}
\usepackage{cases}
\usepackage{tabularx}
\usepackage[top=1.1in, bottom=1.1in, left=1.0in, right=1.0in]{geometry}
\usepackage[utf8]{inputenc}
\usepackage{tikz}
\usepackage{comment}
\usepackage{lineno,hyperref}
\modulolinenumbers[5]
\usepackage{booktabs}
\begin{document}

\title{Semi-implicit Lax-Wendroff kinetic scheme for electron–phonon coupling}
\author[add1]{Jiaming Li}
\ead{780867508@qq.com}
\author[add1]{Hong Liang\corref{cor1}}
\ead{lianghongstefanie@163.com}
\author[add2]{Meng Lian}
\ead{lianmeng@hust.edu.cn}
%\author[add2]{Songze Chen}
%\ead{chensz@tenfong.cn}
\author[add3]{Chuang Zhang\corref{cor1}}
\ead{zhangc26@zju.edu.cn}
\author[add1,add4]{Jiangrong Xu}
\ead{21A3100028@cjlu.edu.cn}
\address[add1]{Department of Physics, Hangzhou Dianzi University, Hangzhou 310018, China}
\address[add2]{School of Physics, Institute for Quantum Science and Engineering and Wuhan National High Magnetic Field Center, Huazhong University of Science and Technology, Wuhan 430074, China}
%{TenFong Technology Company, Nanshan Zhiyuan, No. 1001, Xueyuan Avenue, Taoyuan Street, Nanshan District, Shenzhen, China}
\address[add3]{College of Energy Engineering, Zhejiang University, Hangzhou 310027, China}
\address[add4]{College of Energy Environment and Safety Engineering, China Jiliang University, Hangzhou 310018, China}
\cortext[cor1]{Corresponding author}
\date{\today}

% \title{Semi-implicit Lax-Wendroff kinetic scheme for electron–phonon coupling}
% \author{Jiaming Li}
% %\email{780867508@qq.com}
% \affiliation{Department of Physics, School of Sciences, Hangzhou Dianzi University, Hangzhou 310018, China}
% \author{Hong Liang}
% \email{Corresponding author: lianghongstefanie@163.com}
% \affiliation{Department of Physics, School of Sciences, Hangzhou Dianzi University, Hangzhou 310018, China}
% % \author{Songze Chen}
% % %\email{chensz@tenfong.cn}
% % \affiliation{TenFong Technology Company, Nanshan Zhiyuan, No. 1001, Xueyuan Avenue, Taoyuan Street, Nanshan District, Shenzhen, China}
% \author{Meng Lian}
% % \email{lianmeng@hust.edu.cn}
% \affiliation{School of Physics, Institute for Quantum Science and Engineering and Wuhan National High Magnetic Field Center, Huazhong University of Science and Technology, Wuhan 430074, China}
% \author{Chuang Zhang}
% \email{Corresponding author: zhangc26@zju.edu.cn}
% \affiliation{College of Energy Engineering, Zhejiang University, Hangzhou 310027, China}
% \author{Jiangrong Xu}
% %\email{naxy@cjlu.edu.cn}
% \affiliation{Department of Physics, School of Sciences, Hangzhou Dianzi University, Hangzhou 310018, China}
% \affiliation{College of Energy Environment and Safety Engineering, China Jiliang University, Hangzhou 310018, China}
% \date{\today}

\begin{abstract}

A semi-implicit Lax-Wendroff scheme is developed for electron-phonon coupling process in metals based on the two-temperature kinetic equations. 
The core of this method is to integrate the evolution information of physical equations into the numerical modeling process, which leads to that the time step or cell size is not limited by the relaxation time and mean free path.
Specifically, the finite difference method is used to solve the kinetic model again when reconstructing the interfacial distribution function, through which the particle migration, scattering and electron-phonon coupling processes are coupled together within a single time step.
Numerical tests demonstrate that this method could efficiently capture electron-phonon coupling or heat conduction processes from the ballistic to diffusive regimes. 
It provides a new tool for describing electron–phonon coupling or thermal management in microelectronic devices. 

\end{abstract}

\begin{keyword}
Electron–phonon coupling \sep Multi-scale heat conduction \sep Boltzmann transport equation  \sep Semi-implicit Lax-Wendroff kinetic scheme
\end{keyword}

\maketitle

\section{Introduction}

As the Moore's Law of chips gradually approaches its physical limit~\cite{waldrop_chips_2016,nsrEVIEW_2024} and breakthroughs in packaging or laser technology prompt a transition from the millimeter or second scales to the nanometer or ultra-fast time scales, non-equilibrium thermal transport has become a key frontier foundation in the industrialization process~\cite{TCAD_application_intel_2021_review,kaviany_2008,pop2004analytic,APLnonthermal2020,pop_energy_2010,review_cby_2025}.
At the micro/nano scales, the mean free path or relaxation time of thermal carriers (e.g., electron and phonon) is comparable to the characteristic length or time scale of the system and the classical Fourier heat conduction theory fails, which necessitates a fundamental shift toward characterizing non-equilibrium transport and scattering of thermal carriers~\cite{ChenG05Oxford,ZimanJM60phonons} as the primary determinant of device performance and lifespan. 
Consequently, understanding these transport and interactions mechanisms is the definitive bottleneck for advancing thermal management in next-generation electronics and precision laser machining~\cite{RevModPhys.89.015003,PhysRevB.97.054310,PhysRevB.98.134309,PhysRevB.93.125432}.

Many theoretical models have been developed in the past to describe the fundamental electron-phonon coupling process~\cite{PhysRevLett.59.1460,PhysRevB.74.024301,PhysRevLett.58.1680,PhysRevB.45.5079,TTM_EP_AIPX2022,PhysRevX.6.021003,QIU19942789}.
The two-temperature model (TTM) proposed by Anisimov et al.~\cite{anisimov1974electron}, which describes coupled electron and phonon temperature fields through coupled diffusion equations, has become a useful and simple tool for scenarios like ultra-fast laser heating. 
But the limitations of traditional TTM become increasingly apparent for complex material systems.
It neglects the thermal non-equilibrium between optical and acoustic phonon~\cite{Distinguishing_AdvSci_2020,exp_photon_excitation2021,nano_letters_2017_non_equilibrium}. 
To address this, the non-thermal lattice model~\cite{PhysRevX.6.021003} or multi-temperature model~\cite{PhysRevB.98.134309} was developed, which more precisely describes energy relaxation pathways by establishing separate energy equations for electrons, optical phonon, and acoustic phonon. 
However, both the TTM and MTM rely on diffusion assumption, which neglects non-instantaneous nature of energy transfer~\cite{RevModPhysJoseph89} and fails to capture ballistic transport, nonlocal or nonlinear effects.
Compared to the macroscopic models, the Boltzmann transport equation (BTE) provides a more fundamental theoretical framework for electron–phonon coupling~\cite{IJHMT_2006_EPcoupling_BTE,JHTelectron-phonon2009,PhysRevB.103.125412,PhysRevResearch.3.023072,JAP2016EP_review}.
The phonon/electron BTE naturally encompass carrier scattering, transport and coupling processes through distribution function evolution, enabling a unified description of thermal transport behavior from ballistic to diffusive regimes.
This kinetic model provides a detailed description of non-equilibrium distributions, captures size effects, boundary scattering and self-consistently couples multi-carrier transport.
However, the BTE is a high-dimensional integral-differential equation, posing significant challenges for analytical or numerical solutions.

Over the past decades, numerous numerical methods for solving the BTE have been developed~\cite{ZHANG2025101893,ZHANG2024123379,guo_progress_DUGKS}, such as the lattice Boltzmann method~\cite{JAP2016EP_review,RevModPhys.74.1203}, explicit discrete ordinate method~\cite{PhysRevB.103.125412,MIAO2023123538} and Monte Carlo method~\cite{MUTHUKUNNILJOSEPH2022107742}. 
These methods have played significant and successful roles at different scales. 
However, it is difficult for them to maintain extremely high computational efficiency and accuracy for all Knudsen numbers due to the introduction of some approximations or assumptions in numerical modeling process~\cite{JinS10AP,PhysRevE.107.025301}. 
The lattice Boltzmann method is unable to accurately describe the heat transfer phenomena in the highly non-equilibrium or ballistic regime due to the limited discrete points in the momentum space~\cite{RevModPhys.74.1203}. 
The latter two methods decouple the particle migration and scattering on a numerical time step scale. 
Therefore, the time step is limited by the relaxation time, making them difficult to efficiently simulate unsteady thermal transport problems with extremely small Knudsen numbers~\cite{DSMC_book_1994,ZHANG2024123379,guo_progress_DUGKS}.

Recently, a semi-implicit Lax-Wendroff kinetic scheme has been developed for solving BTE~\cite{PENG202572}, demonstrating its ability to capture transitions from ballistic to diffusive transient heat transfer and its independence from relaxation time constraints~\cite{csz_Lax_Wendroff2022}. 
Given the excellent performance of this method in a single physical field~\cite{csz_Lax_Wendroff2022,PENG202572}, this study further expands it and applies it to the multi-scale heat conduction problem with electron-phonon coupling.
The rest of the paper is organized as follows. In Sec.~\ref{sec:bte}, the kinetic model is introduced as well as the semi-implicit Lax-Wendroff scheme. 
Numerical simulations and tests are shown in Sec.~\ref{sec:results}.
Conclusions are made in Sec.~\ref{sec:conclusion}.

\section{Model and Numerical Algorithm}
\label{sec:bte}

\subsection{Electron and Phonon BTE}

The Boltzmann transport equation for electron-phonon coupling is expressed as follows~\cite{ChenG05Oxford,ZHANG2025101893,ZHANG2024123379,IJHMT_2006_EPcoupling_BTE,JHTelectron-phonon2009,PhysRevB.103.125412,PhysRevResearch.3.023072,JAP2016EP_review},
\begin{align}
\frac{\partial f_{e}}{\partial t}+\bm{v_e} \cdot \nabla_{\bm{x}} f_{e}=\frac{1}{\tau_e } \left(  {f}^{eq}_{e} - f_{e}\right)-\frac{G}{4 \pi}(T_{e}-T_{p}),\label{1}
\end{align}
\begin{align}
\frac{\partial f_{p}}{\partial t}+\bm{v_p} \cdot \nabla_{\bm{x}} f_{p} =\frac{1}{\tau_p } \left(  {f}^{eq}_{p} - f_{p}\right)+\frac{G}{4 \pi}(T_{e}-T_{p}).\label{2}
\end{align}
where the subscripts $e$ and $p$ represent electron and phonon, respectively. 
These subscripts will be consistently adopted throughout the remainder of this paper. 
The electron and phonon energy density distribution functions are given by $f_{e}=f_{e}(\bm{x},\bm{v_e},t)$ and $f_{p}=f_{p}(\bm{x},\bm{v_p},t)$, respectively, which depend on the spatial position $\bm{x}$, time $t$ and group velocity $ \bm{v_e} = \left|\bm{ v_e } \right| \bm{s} $ or $ \bm{v_p} = \left|\bm{ v_p } \right| \bm{s} $.
$\bm s=\left( \cos \theta, \sin \theta \cos \varphi, \sin \theta \sin \varphi \right)$ is the unit directional vector with polar angle $\theta$ and azimuthal angle $\varphi$. 
$\tau$ is the relaxation time and the electron-phonon coupling constant is denoted by $G$. 
The equilibrium distribution functions for electron and phonon are given as follows,
\begin{align}
{f}^{eq}_{e}=\frac{C_eT_{e}}{4 \pi}, \quad
{f}^{eq}_{p}=\frac{C_pT_{p}}{4 \pi} .
\label{eq:equilibrium state}
\end{align}
where $C$ is the specific heat and $T$ is the temperature.
Energy conservation is satisfied during the electron-electron, phonon-phonon scattering processes,
\begin{align}
\int \frac{{f}^{eq}_{e}-f_e}{\tau_e} d\Omega= 0,\quad 
\int \frac{{f}^{eq}_{p}-f_p}{\tau_p } d\Omega=0. \label{energy conservation}
\end{align}
where $d\Omega$ represents the integral over the whole solid angle space.
Macroscopic variables including the local energy density $U$, heat flux $\bm{q}$ and temperature $T$ are obtained by taking the moment of distribution function.
\begin{align}
{U}_{e}=\int{f}_{e}d\Omega , \quad   
{q}_{e}=\int{v}_{e}f_ed\Omega,  \quad
{T}_{e}=\frac{1}{{C}_{e}}\int {f}_{e}d\Omega=\frac{{U}_{e}}{{C}_{e}}, \label{Ue}  \\
{U}_{p}=\int{f}_{p}d\Omega , \quad 
{q}_{p}=\int{v}_{p}f_pd\Omega,\quad  
{T}_{p}=\frac{1}{{C}_{p}}\int {f}_{p}d\Omega=\frac{{U}_{p}}{{C}_{p}}. 
\label{Up}
\end{align}

\subsection{Semi-implicit Lax-Wendroff kinetic scheme}

The semi-implicit Lax–Wendroff kinetic scheme~\cite{PENG202572} is introduced for electron–phonon coupling in detail.
Taking a two-dimensional kinetic model as an example for ease of understanding and keeping generality, the discretization process of the present scheme is,
\begin{align}
\frac{\partial f_e}{\partial t}+u_e\frac{\partial f_e}{\partial x}+v_e\frac{\partial f_e}{\partial y}=\frac{1}{\tau_e}\left ({f}^{eq}_{e}-f_{e}\right )-\frac{G}{4 \pi}(T_{e}-T_{p}),\label{7}
\end{align}
\begin{align}
\frac{\partial f_p}{\partial t}+u_p\frac{\partial f_p}{\partial x}+v_p\frac{\partial f_p}{\partial y}=\frac{1}{\tau_p}\left ({f}^{eq}_{p}-{f}_{p}\right )+\frac{G}{4 \pi}(T_{e}-T_{p}),\label{8}
\end{align}
where $u=|v| \cos \theta$ and $v=|v| \sin \theta \cos \varphi$ are the $x-$component and $y-$component of the group velocity, respectively.
Uniform Cartesian grid is used for both the spatial and temporal spaces under the framework of finite volume method, where $(\Delta x,~N_x,~i)$ and $(\Delta y,~N_y,~j)$ are the (cell size, total cell number, index of cell center) in the $x$ and $y$ direction, respectively.
The solid angle space is discretized with the index $k$, too.

Taking an integral of the governing equations over the time interval from $t^n$ to $t^{n+1}=t^{n}+\Delta t$ for a controlled volume leads to,
\begin{align}
\frac{{f}_{e,i,j,k}^{n+1}-{f}_{e,i,j,k}^{n}}{\Delta t}+{u}_{e,k}\frac{\partial {f}_{e,i,j,k}^{n+1/2}}{\partial x}+{v}_{e,k}\frac{\partial {f}_{e,i,j,k}^{n+1/2}}{\partial y}&= \frac{1}{2}\left ( \frac{{f}_{e,i,j,k}^{eq,n+1}-{f}_{e,i,j,k}^{n+1}}{{\tau }_{e}}+\frac{{f}_{e,i,j,k}^{eq,n}-{f}_{e,i,j,k}^{n}}{{\tau }_{e}}\right )\nonumber\\&  -\frac{G}{8 \pi}\left ( {T}_{e,i,j}^{n+1}-{T}_{p,i,j}^{n+1}+{T}_{e,i,j}^{n}-{T}_{p,i,j}^{n}\right ),\label{9}
\end{align}
\begin{align}
\frac{{f}_{p,i,j,k}^{n+1}-{f}_{p,i,j,k}^{n}}{\Delta t}+{u}_{p,k}\frac{\partial {f}_{p,i,j,k}^{n+1/2}}{\partial x}+{v}_{p,k}\frac{\partial {f}_{p,i,j,k}^{n+1/2}}{\partial y}&=\frac{1}{2}\left ( \frac{{f}_{p,i,j,k}^{eq,n+1}-{f}_{p,i,j,k}^{n+1}}{{\tau }_{p}}+\frac{{f}_{p,i,j,k}^{eq,n}-{f}_{p,i,j,k}^{n}}{{\tau }_{p}}\right )\nonumber\\& +\frac{G}{8 \pi}\left ( {T}_{e,i,j}^{n+1}-{T}_{p,i,j}^{n+1}+{T}_{e,i,j}^{n}-{T}_{p,i,j}^{n}\right ),\label{10}
\end{align}
where ${f}_{e,i,j,k}^{n}={f}_{e}({x}_{i},{y}_{j},u_{e,k},v_{e,k},{t}^{n})$, ${f}_{p,i,j,k}^{n}={f}_{p}({x}_{i},{y}_{j},u_{p,k},v_{p,k},{t}^{n})$.
The right-hand sides of the above equation are discretized by the trapezoidal rule and the flux terms are handled via the midpoint rule.          
The spatial gradients of distribution function are 
\begin{align}
\frac{\partial {f}_{i,j,k}^{n+1/2}}{\partial x} =\frac{{f}_{i+1/2,j,k}^{n+1/2}-{f}_{i-1/2,j,k}^{n+1/2}}{\Delta x} ,  \quad  
\frac{\partial {f}_{i,j,k}^{n+1/2}}{\partial y} =\frac{{f}_{i,j+1/2,k}^{n+1/2}-{f}_{i,j-1/2,k}^{n+1/2}}{\Delta y},\label{11}
\end{align}
where $(i \pm 1/2,j)$ or $(i,j \pm 1/2)$ represents the indexes of cell interfaces connected to cell center $(i,j)$ in the $x-$ and $y-$ direction, respectively.

Reformulating Eqs.(\ref{9},\ref{10}),
\begin{align}
{f}_{e,i,j,k}^{n+1} & =\frac{{\tau }_{e}}{{\tau }_{e}+h}{f}_{e,i,j,k}^{n}+\frac{h}{{\tau }_{e}+h}{f}_{e,i,j,k}^{eq,n+1}+\frac{h}{{\tau }_{e}+h}\left ({f}_{e,i,j,k}^{eq,n} -{f}_{e,i,j,k}^{n}\right )\nonumber\\&   -\frac{2h{\tau}_{e}}{{\tau }_{e}+h}\left ( {u}_{e,k}\frac{{f}_{e,i+1/2,j,k}^{n+1/2}-{f}_{e,i-1/2,j,k}^{n+1/2}}{\Delta x}+{v}_{e,k}\frac{{f}_{e,i,j+1/2,k}^{n+1/2}-{f}_{e,i,j-1/2,k}^{n+1/2}}{\Delta y}\right )\nonumber\\ & -\frac{h{\tau}_{e}}{{\tau }_{e}+h}\frac{G}{4 \pi}\left ({T}_{e,i,j}^{n+1}-{T}_{p,i,j}^{n+1}+{T}_{e,i,j}^{n}-{T}_{p,i,j}^{n}\right ),\label{(n+1)distribution function of e}
\end{align}
\begin{align}
{f}_{p,i,j,k}^{n+1} & =\frac{{\tau }_{p}}{{\tau }_{p}+h}{f}_{p,i,j,k}^{n}+\frac{h}{{\tau }_{p}+h}{f}_{p,i,j,k}^{eq,n+1}+\frac{h}{{\tau }_{p}+h}\left ({f}_{p,i,j,k}^{eq,n} -{f}_{p,i,j,k}^{n}\right )\nonumber\\&   -\frac{2h{\tau}_{p}}{{\tau }_{p}+h}\left ( {u}_{p,k}\frac{{f}_{p,i+1/2,j,k}^{n+1/2}-{f}_{p,i-1/2,j,k}^{n+1/2}}{\Delta x}+{v}_{p,k}\frac{{f}_{p,i,j+1/2,k}^{n+1/2}-{f}_{p,i,j-1/2,k}^{n+1/2}}{\Delta y}\right )\nonumber\\ & +\frac{h{\tau}_{p}}{{\tau }_{p}+h}\frac{G}{4 \pi}\left ({T}_{e,i,j}^{n+1}-{T}_{p,i,j}^{n+1}+{T}_{e,i,j}^{n}-{T}_{p,i,j}^{n}\right ),\label{(n+1)distribution function of p}
\end{align}
where $h=\Delta t/2$.
Taking the moment of Eqs.(\ref{9},\ref{10}) leads to the macroscopic governing equation at the cell center,
\begin{align}
{U}_{e,i,j}^{n+1}& ={U}_{e,i,j}^{n}-\Delta t\sum_{k}\left ({u}_{e,k}\frac{{f}_{e,i+1/2,j,k}^{n+1/2}-{f}_{e,i-1/2,j,k}^{n+1/2}}{\Delta x}+{v}_{e,k}\frac{{f}_{e,i,j+1/2,k}^{n+1/2}-{f}_{e,i,j-1/2,k}^{n+1/2}}{\Delta y} \right ){\phi }_{k}\nonumber\\ & - hG\left ( {T}_{e,i,j}^{n+1}-{T}_{p,i,j}^{n+1}+{T}_{e,i,j}^{n}-{T}_{p,i,j}^{n}\right ), \label{(n+1)Ue}
\end{align}
\begin{align}
{U}_{p,i,j}^{n+1} & ={U}_{p,i,j}^{n}-\Delta t\sum_{k}\left ({u}_{p,k}\frac{{f}_{p,i+1/2,j,k}^{n+1/2}-{f}_{p,i-1/2,j,k}^{n+1/2}}{\Delta x}+{v}_{p,k}\frac{{f}_{p,i,j+1/2,k}^{n+1/2}-{f}_{p,i,j-1/2,k}^{n+1/2}}{\Delta y} \right ){\phi }_{k}\nonumber\\ & + hG\left ( {T}_{e,i,j}^{n+1}-{T}_{p,i,j}^{n+1}+{T}_{e,i,j}^{n}-{T}_{p,i,j}^{n}\right ),\label{(n+1)Up}
\end{align}
where $\sum_{k}$ is the numerical quadrature over the discretized solid angle space and ${\phi }_{k}$ is the associated weight.

Based on Eqs.(\ref{eq:equilibrium state},\ref{Ue},\ref{Up}), the equilibrium state is completely determined by the local temperature or energy density. 
Combining the above two equations (\ref{(n+1)Ue},\ref{(n+1)Up}) yields a system of two linear equations in two variables concerning electron and phonon temperatures at the next time step.
Therefore, the key step is calculating the unknown electron and phonon distribution functions at the cell interfaces  at the half-time step ($t^{n+1/2}$).
Taking an integral of the BTE from $t^{n}$ to $t^{n+1/2}=t^{n}+h$ at the cell interfaces $(x_{i+1/2},y_j)$ and $(x_i,y_{j+1/2})$ leads to,
\begin{align}
\frac{{f}_{e,i+1/2,j,k}^{n+1/2}-{f}_{e,i+1/2,j,k}^{n}}{h}+{u}_{e,k}\frac{\partial {f}_{e,i+1/2,j,k}^{n}}{\partial x}+{v}_{e,k}\frac{\partial {f}_{e,i+1/2,j,k}^{n}}{\partial y} &= \frac{1}{{\tau }_{e}}\left ( {f}_{e,i+1/2,j,k}^{eq,n+1/2}-{f}_{e,i+1/2,j,k}^{n+1/2}\right )\nonumber\\ &-\frac{G}{4 \pi}\left ({T}_{e,i+1/2,j}^{n+1/2}- {T}_{p,i+1/2,j}^{n+1/2}\right ),\label{22}
\end{align}
\begin{align}
\frac{{f}_{e,i,j+1/2,k}^{n+1/2}-{f}_{e,i,j+1/2,k}^{n}}{h}+{u}_{e,k}\frac{\partial {f}_{e,i,j+1/2,k}^{n}}{\partial x}+{v}_{e,k}\frac{\partial {f}_{e,i,j+1/2,k}^{n}}{\partial y} &= \frac{1}{{\tau }_{e}}\left ( {f}_{e,i,j+1/2,k}^{eq,n+1/2}-{f}_{e,i,j+1/2,k}^{n+1/2}\right )\nonumber\\ &-\frac{G}{4 \pi}\left ({T}_{e,i,j+1/2}^{n+1/2}- {T}_{p,i,j+1/2}^{n+1/2}\right ),\label{23}
\end{align}
\begin{align}
\frac{{f}_{p,i+1/2,j,k}^{n+1/2}-{f}_{p,i+1/2,j,k}^{n}}{h}+{u}_{p,k}\frac{\partial {f}_{p,i+1/2,j,k}^{n}}{\partial x}+{v}_{p,k}\frac{\partial {f}_{p,i+1/2,j,k}^{n}}{\partial y} &= \frac{1}{{\tau }_{p}}\left ( {f}_{p,i+1/2,j,k}^{eq,n+1/2}-{f}_{p,i+1/2,j,k}^{n+1/2}\right )\nonumber\\ &+\frac{G}{4 \pi}\left ({T}_{e,i+1/2,j}^{n+1/2}- {T}_{p,i+1/2,j}^{n+1/2}\right ),\label{24}
\end{align}
\begin{align}
\frac{{f}_{p,i,j+1/2,k}^{n+1/2}-{f}_{p,i,j+1/2,k}^{n}}{h}+{u}_{p,k}\frac{\partial {f}_{p,i,j+1/2,k}^{n}}{\partial x}+{v}_{p,k}\frac{\partial {f}_{p,i,j+1/2,k}^{n}}{\partial y} &= \frac{1}{{\tau }_{p}}\left ( {f}_{p,i,j+1/2,k}^{eq,n+1/2}-{f}_{p,i,j+1/2,k}^{n+1/2}\right )\nonumber\\ &+\frac{G}{4 \pi}\left ({T}_{e,i,j+1/2}^{n+1/2}- {T}_{p,i,j+1/2}^{n+1/2}\right ),\label{25}
\end{align}
where the right-hand sides of the above equation are discretized by the forward-Euler rule and the flux terms are handled via the backward-Euler rule.        
Making a transformation of above equations leads to,
\begin{align}
{f}_{e,i+1/2,j,k}^{n+1/2}   &=\frac{{\tau }_{e}}{{\tau }_{e}+h}{f}_{e,i+1/2,j,k}^{n}-\frac{{\tau }_{e}h}{{\tau }_{e}+h}\left ( {u}_{e,k}\frac{\partial {f}_{e,i+1/2,j,k}^{n}}{\partial x}+{v}_{e,k}\frac{\partial {f}_{e,i+1/2,j,k}^{n}}{\partial y}\right )+\frac{h}{{\tau }_{e}+h}{f}_{e,i+1/2,j,k}^{eq,n+1/2}\nonumber\\ &-\frac{{\tau }_{e}h}{{\tau }_{e}+h}\frac{G}{4 \pi}({T}_{e,i+1/2,j}^{n+1/2}-{T}_{p,i+1/2,j}^{n+1/2}),\label{distribution function of e ihalf}
\end{align}
\begin{align}
{f}_{e,i,j+1/2,k}^{n+1/2}  &=\frac{{\tau }_{e}}{{\tau }_{e}+h}{f}_{e,i,j+1/2,k}^{n}-\frac{{\tau }_{e}h}{{\tau }_{e}+h}\left ( {u}_{e,k}\frac{\partial {f}_{e,i,j+1/2,k}^{n}}{\partial x}+{v}_{e,k}\frac{\partial {f}_{e,i,j+1/2,k}^{n}}{\partial y}\right )+\frac{h}{{\tau }_{e}+h}{f}_{e,i,j+1/2,k}^{eq,n+1/2}\nonumber\\ &-\frac{{\tau }_{e}h}{{\tau }_{e}+h}\frac{G}{4 \pi}({T}_{e,i,j+1/2}^{n+1/2}-{T}_{p,i,j+1/2}^{n+1/2}),\label{distribution function of e jhalf}
\end{align}
\begin{align}
{f}_{p,i+1/2,j,k}^{n+1/2}   &=\frac{{\tau }_{p}}{{\tau }_{p}+h}{f}_{p,i+1/2,j,k}^{n}-\frac{{\tau }_{p}h}{{\tau }_{p}+h}\left ( {u}_{p,k}\frac{\partial {f}_{p,i+1/2,j,k}^{n}}{\partial x}+{v}_{p,k}\frac{\partial {f}_{p,i+1/2,j,k}^{n}}{\partial y}\right )+\frac{h}{{\tau }_{p}+h}{f}_{p,i+1/2,j,k}^{eq,n+1/2}\nonumber\\ &+\frac{{\tau }_{p}h}{{\tau }_{p}+h}\frac{G}{4 \pi}({T}_{e,i+1/2,j}^{n+1/2}-{T}_{p,i+1/2,j}^{n+1/2}),\label{distribution function of p ihalf}
\end{align}
\begin{align}
{f}_{p,i,j+1/2,k}^{n+1/2}  &=\frac{{\tau }_{p}}{{\tau }_{p}+h}{f}_{p,i,j+1/2,k}^{n}-\frac{{\tau }_{p}h}{{\tau }_{p}+h}\left ( {u}_{p,k}\frac{\partial {f}_{p,i,j+1/2,k}^{n}}{\partial x}+{v}_{p,k}\frac{\partial {f}_{p,i,j+1/2,k}^{n}}{\partial y}\right )+\frac{h}{{\tau }_{p}+h}{f}_{p,i,j+1/2,k}^{eq,n+1/2}\nonumber\\ &+\frac{{\tau }_{p}h}{{\tau }_{p}+h}\frac{G}{4 \pi}({T}_{e,i,j+1/2}^{n+1/2}-{T}_{p,i,j+1/2}^{n+1/2}),\label{distribution function of p jhalf}
\end{align}
Taking the moment of Eqs.(\ref{(n+1/2)Ue_ihalf},\ref{(n+1/2)Ue_jhalf},\ref{(n+1/2)Up_ihalf},\ref{(n+1/2)Up_jhalf}) leads to the macroscopic governing equation at the cell interface,
\begin{align}
{U}_{e,i+1/2,j}^{n+1/2} ={U}_{e,i+1/2,j}^{n}-h\sum_{k}\left ( {u}_{e,k}\frac{\partial {f}_{e,i+1/2,j,k}^{n}}{\partial x}+{v}_{e,k}\frac{\partial {f}_{e,i+1/2,j,k}^{n}}{\partial y}\right){\phi }_{k}- hG\left ( {T}_{e,i+1/2,j}^{n+1/2}-{T}_{p,i+1/2,j}^{n+1/2}\right ),\label{(n+1/2)Ue_ihalf}
\end{align}
\begin{align}
{U}_{e,i,j+1/2}^{n+1/2} ={U}_{e,i,j+1/2}^{n}-h\sum_{k}\left ( {u}_{e,k}\frac{\partial {f}_{e,i,j+1/2,k}^{n}}{\partial x}+{v}_{e,k}\frac{\partial {f}_{e,i,j+1/2,k}^{n}}{\partial y}\right){\phi }_{k}- hG\left ( {T}_{e,i,j+1/2}^{n+1/2}-{T}_{p,i,j+1/2}^{n+1/2}\right ),\label{(n+1/2)Ue_jhalf}
\end{align}
\begin{align}
{U}_{p,i+1/2,j}^{n+1/2} ={U}_{p,i+1/2,j}^{n}-h\sum_{k}\left ( {u}_{p,k}\frac{\partial {f}_{p,i+1/2,j,k}^{n}}{\partial x}+{v}_{p,k}\frac{\partial {f}_{p,i+1/2,j,k}^{n}}{\partial y}\right){\phi }_{k}+ hG\left ( {T}_{e,i+1/2,j}^{n+1/2}-{T}_{p,i+1/2,j}^{n+1/2}\right ),\label{(n+1/2)Up_ihalf}
\end{align}
\begin{align}
{U}_{p,i,j+1/2}^{n+1/2} ={U}_{p,i,j+1/2}^{n}-h\sum_{k}\left ( {u}_{p,k}\frac{\partial {f}_{p,i,j+1/2,k}^{n}}{\partial x}+{v}_{p,k}\frac{\partial {f}_{p,i,j+1/2,k}^{n}}{\partial y}\right){\phi }_{k}+ hG\left ( {T}_{e,i,j+1/2}^{n+1/2}-{T}_{p,i,j+1/2}^{n+1/2}\right ).\label{(n+1/2)Up_jhalf}
\end{align}
In order to obtain $U^{n+1/2}$ or $T^{n+1/2}$, it is inevitable is to calculate the interfacial distribution function and its spatial gradients at the $n-$time step ($t^{n}$) based on above four equations.
Fortunately, the detailed calculation formula of interfacial distribution function and its spatial gradients has been written in the previous paper~\cite{PENG202572}.

The procedure of the present scheme can be summarized as follows:
\begin{enumerate}
    \item At $t^{n}$, the interfacial distribution function and its associated spatial gradients are calculated by a second-order interpolation method, as done in a previous work~\cite{PENG202572}. 
    \item At $t^{n+1/2}$, update the macroscopic fields at the cell interface based on Eqs. (\ref{(n+1/2)Ue_ihalf},\ref{(n+1/2)Ue_jhalf},\ref{(n+1/2)Up_ihalf},\ref{(n+1/2)Up_jhalf},\ref{Ue},\ref{Up}). Then, update the equilibrium distribution function of phonon/electron based on Eq. (\ref{eq:equilibrium state}).
    \item At $t^{n+1/2}$, update the electron/phonon distribution function at the cell interface based on Eqs. (\ref{distribution function of e ihalf},\ref{distribution function of e jhalf},\ref{distribution function of p ihalf},\ref{distribution function of p jhalf}).
    \item At $t^{n+1}$, update the macroscopic fields at the cell center based on Eqs. (\ref{(n+1)Ue},\ref{(n+1)Up},\ref{Ue},\ref{Up}). Then, update the equilibrium distribution function of phonon/electron based on Eq. (\ref{eq:equilibrium state}).
    \item At $t^{n+1}$, update phonon/electron distribution function at the cell center based on Eqs. (\ref{(n+1)distribution function of e},\ref{(n+1)distribution function of p}).
\end{enumerate}

\section{Numerical tests}
\label{sec:results}

The performance of the present scheme for electron-phonon coupling is validated by numerical tests in this part. 
Numerical results are compared with those predicted by typical two-temperature models (TTM), explicit discrete ordinate method (DOM) and discrete unified gas kinetic (DUGKS) in previous references~\cite{ZHANG2024123379}.
In quasi-1D simulations, the Gauss–Legendre quadrature is used to discrete $\cos \theta \in [-1,1]$ into $N_{\theta}$ points.
In quasi-2D or 3D simulations, the Gauss–Legendre quadrature is also used to discretize the azimuthal angle $ \varphi \in [0,\pi]$ into $N_{\varphi} /2$ points due to symmetry.
All simulations were executed using a single core of AMD Ryzen 7 5800H with Radeon Graphics.
The Knudsen number is defined as the ratio of the mean free path $\lambda=|\bm{v}| \tau $ to the characteristic length of the system, i.e., $\text{Kn}= \lambda / L $.
The physical time step $\Delta t$ is 
\begin{align}
\Delta t = \text{CFL} \times \frac{\left \{ \Delta x,\Delta y\right \}_{min}}{\left \{ |{v_e}|,|{v_p}|\right \}_{max}} 
\end{align}
where $0< \text{CFL} <1$ is the Courant–Friedrichs–Lewy number.
Without special statements, CFL=$0.40$. 

\subsection{Quasi-1D heat conduction}

Quasi-1D cross-plane heat conduction in Au metal films with varying thicknesses is investigated.
Isothermal boundary conditions are implemented for the left and right ends of the film with constant temperatures ${T}_{L}={T}_{0}+\Delta T$ and ${T}_{R}={T}_{0}-\Delta T$, respectively.
Detailed thermophysical parameters of Au metals are shown below: electron mean free path ${\lambda }_{e}$=33 nm, phonon mean free path ${\lambda}_{p}$=1.5 nm, electron relaxation time ${\tau }_{e}$=0.0243 ps and phonon relaxation time ${\tau }_{p}$=0.679 ps.
$40$ discretized cells and ${N}_{\theta }$=48 discretized angles are used.
The convergence is reach when 
\begin{align}
\epsilon =\frac{\sqrt{\textstyle\sum_{i}\left ({T}_{i}^{n}-{T}_{i}^{n+1} \right )^2}}{\sqrt{\textstyle\sum_{i}\left (\Delta T \right )^2}}<{10}^{-8} 
\end{align}

\begin{figure}[htb]
    \centering
    \subfloat[$L$=10 nm, $\Delta x$=0.25 nm]{\includegraphics[width=0.33\textwidth]{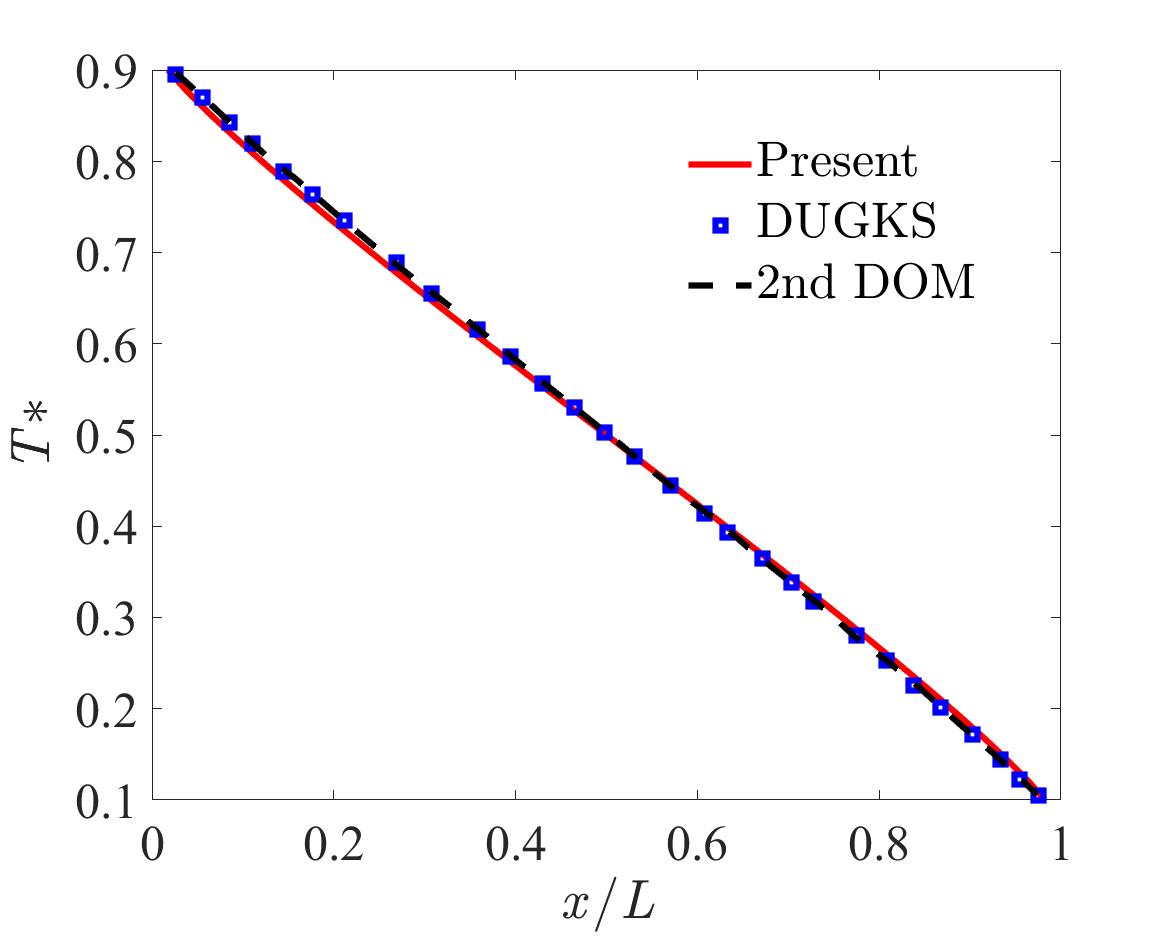}}~~
    \subfloat[$L$=100 nm, $\Delta x$=2.5 nm]{\includegraphics[width=0.33\textwidth]{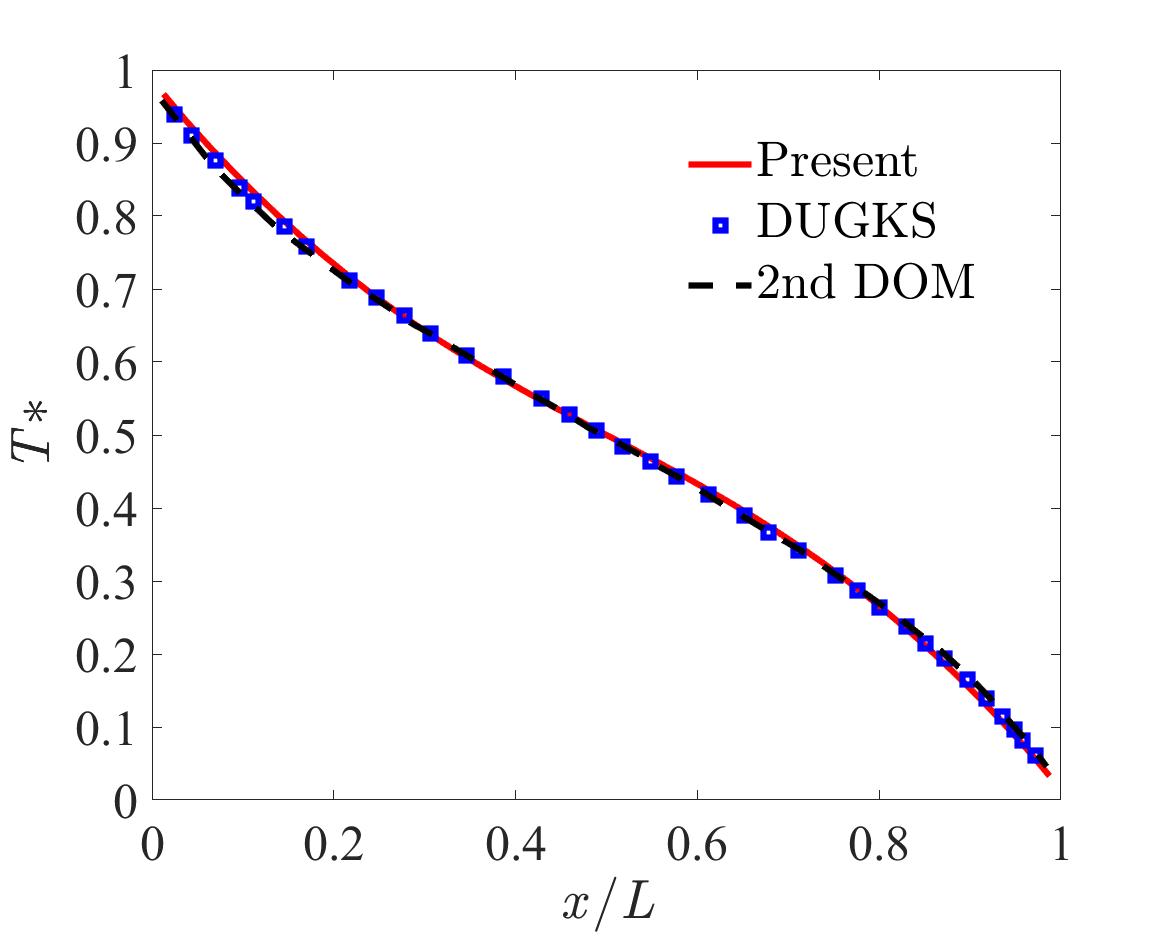}}~~
    \subfloat[$L$=10 $\mu$m$, \Delta x$=250 nm]{\includegraphics[width=0.33\textwidth]{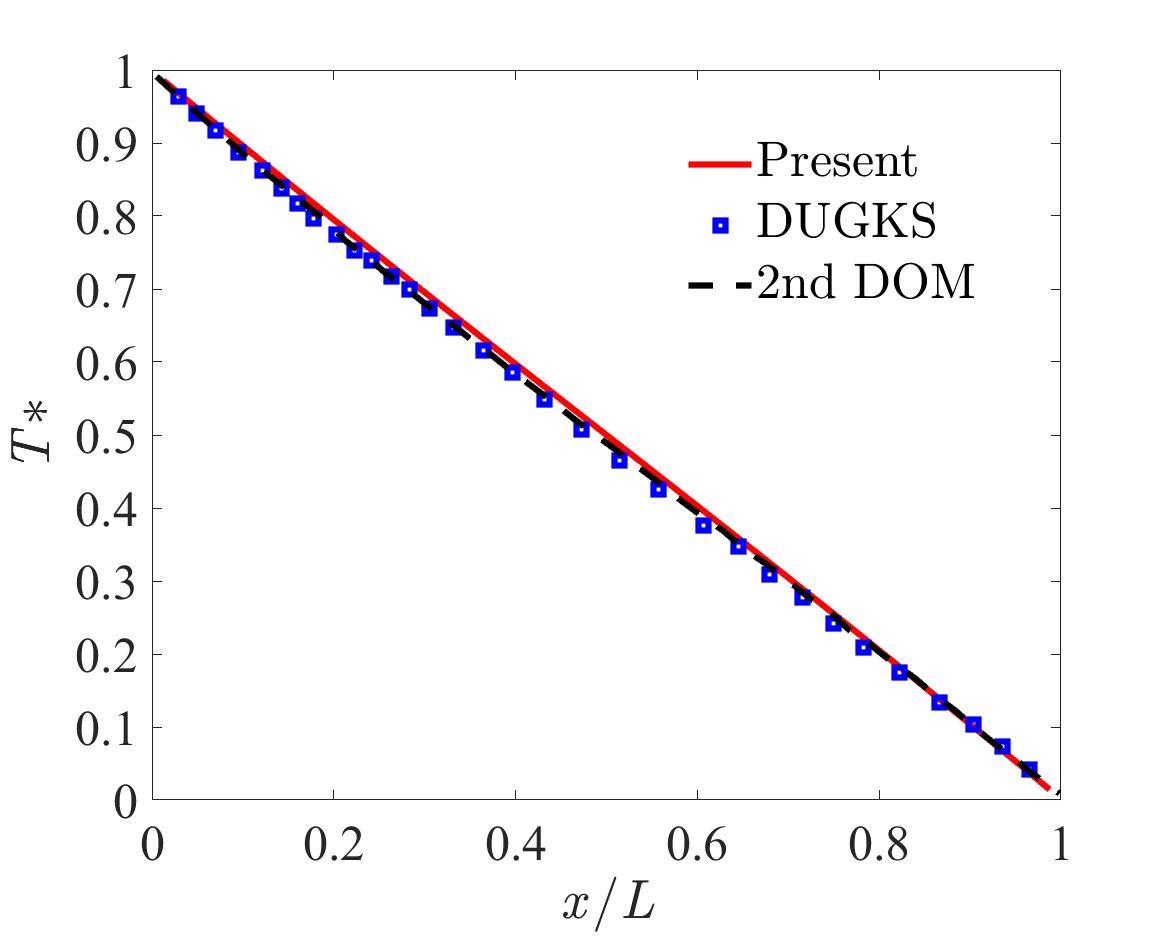}}
    \caption{
    Spatial distributions of the phonon temperatures with different thickness $L$, where $T^* = (T - T_R) / (T_L - T_R)$.
   }
    \label{fig1}
\end{figure}

The spatial distributions of the phonon temperature at various thicknesses are plotted in~\cref{fig1}. 
A transition of the phonon temperature distribution from non-linear to near-linear is observed as the film thickness is increased from the nanometer to the micrometer scale, indicating a shift in the dominant heat conduction mechanism from the ballistic to diffusive. 
Specifically, pronounced nonlinearity is exhibited by the temperature curve when $L$=100 nm, which signifies the presence of significant ballistic transport effects in energy carriers. 
In contrast, a nearly linear distribution is approached by the temperature profile at 10 $\mu$m, which is consistent with the description of Fourier's diffusion law, demonstrating the dominance of diffusion mechanisms at this scale.
The reliable consistency of the present method with previous methods across a broad range of scales, from ballistic to diffusive regime, is demonstrated, confirming its suitability for multiscale heat conduction analysis.

\subsection{Transient thermal grating}

Transient thermal grating technique is an advanced approach for measuring the thermal properties of nanomaterials~\cite{collins_non-diffusive_2013,TTG_metals_2011_JAP,sivan_ultrafast_2020}.
A periodic temperature distribution, for instance of a sinusoidal or cosine shape, 
\begin{align}
T_{e}(x,t=0)=T_{ref}+\Delta T \sin (2\pi x/L),  \quad  T_{p}(x,t=0)=T_{ref}. 
\end{align}
is achieved on the sample surface by the fundamental principle wherein optical interference fringes are generated by two coherent pump beams, where $\Delta T$ is the temperature amplitude, $L$ is the spatial grating period length and the laser heating system initially only heats the electronic system. 
To simulate transient thermal phenomenon, $20$-$100$ discretized cells and ${N}_{\theta }$=$8$-$100$ discretized angles are used. 
Periodic boundary conditions were applied to the left and right boundaries of single spatial grating period. 
Physical parameters of BTE in Au metals at 300 K and 25 K were obtained from previous reference.
\begin{figure}[h]
    \centering
    \subfloat[$L$=100 nm, $\Delta x$=2.5 nm]{\label{fig3a}\includegraphics[width=0.40\textwidth]{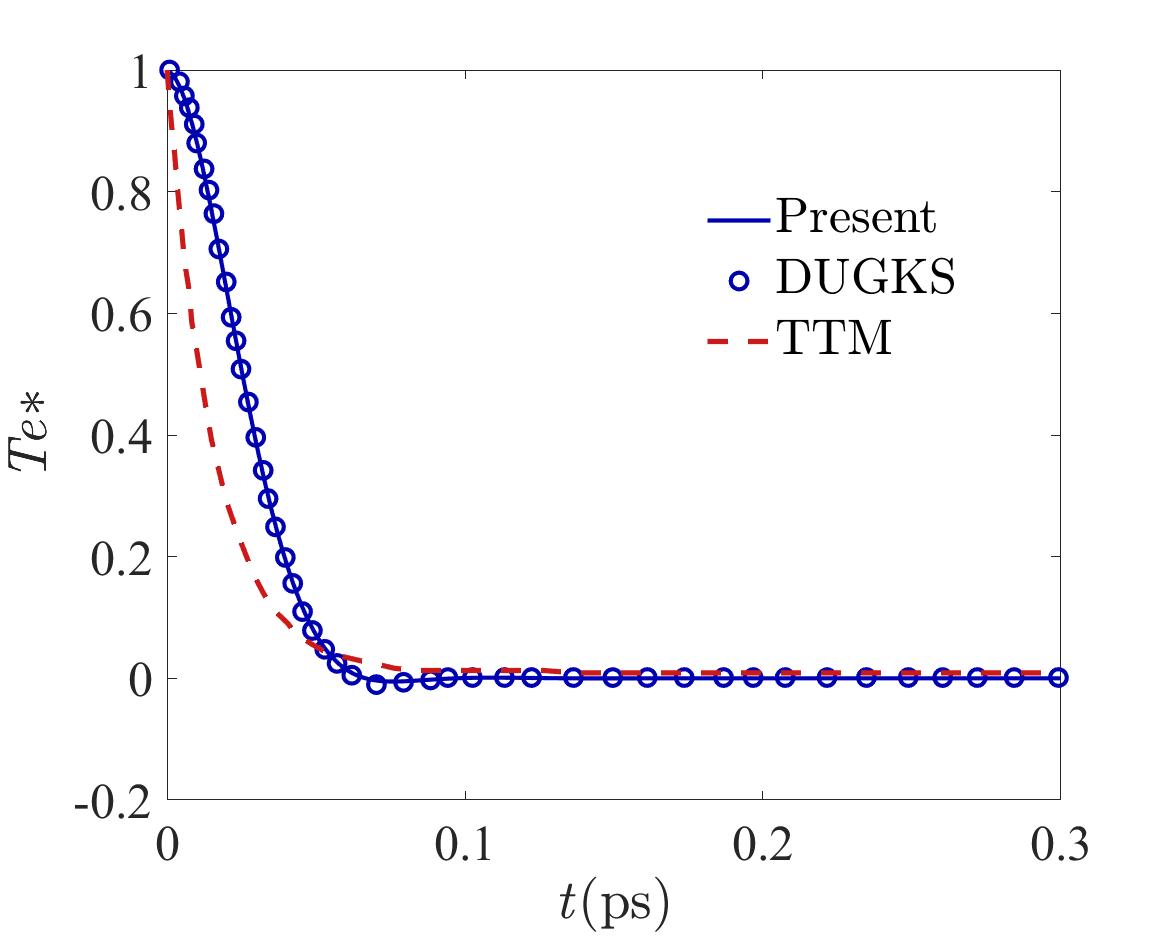}}~~ 
    \subfloat[$L$=1 $\mu$m, $\Delta x$=25 nm]{\label{fig3b}\includegraphics[width=0.40\textwidth]{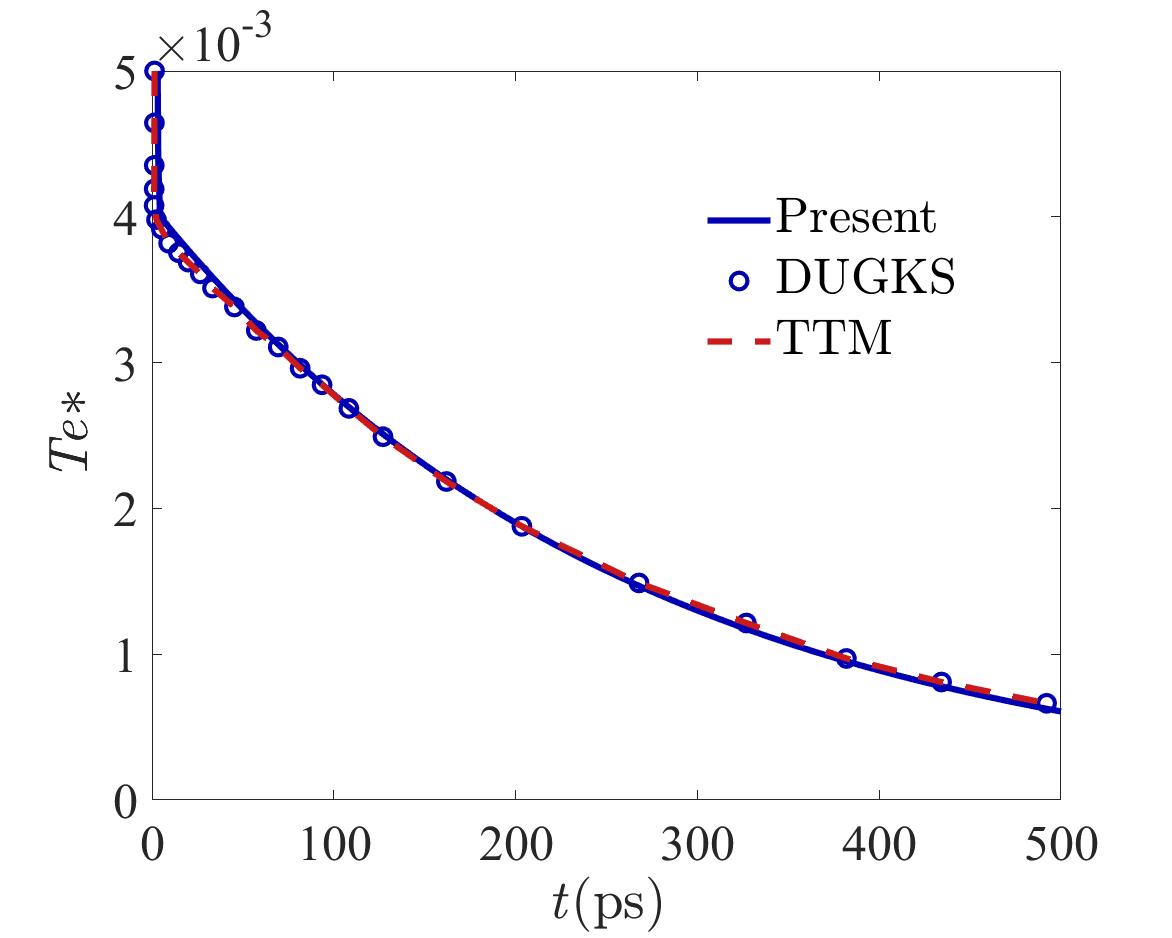}}\\ 
    \subfloat[$L$=100 nm, $\Delta x$=2.5 nm]{\label{fig3c}\includegraphics[width=0.40\textwidth]{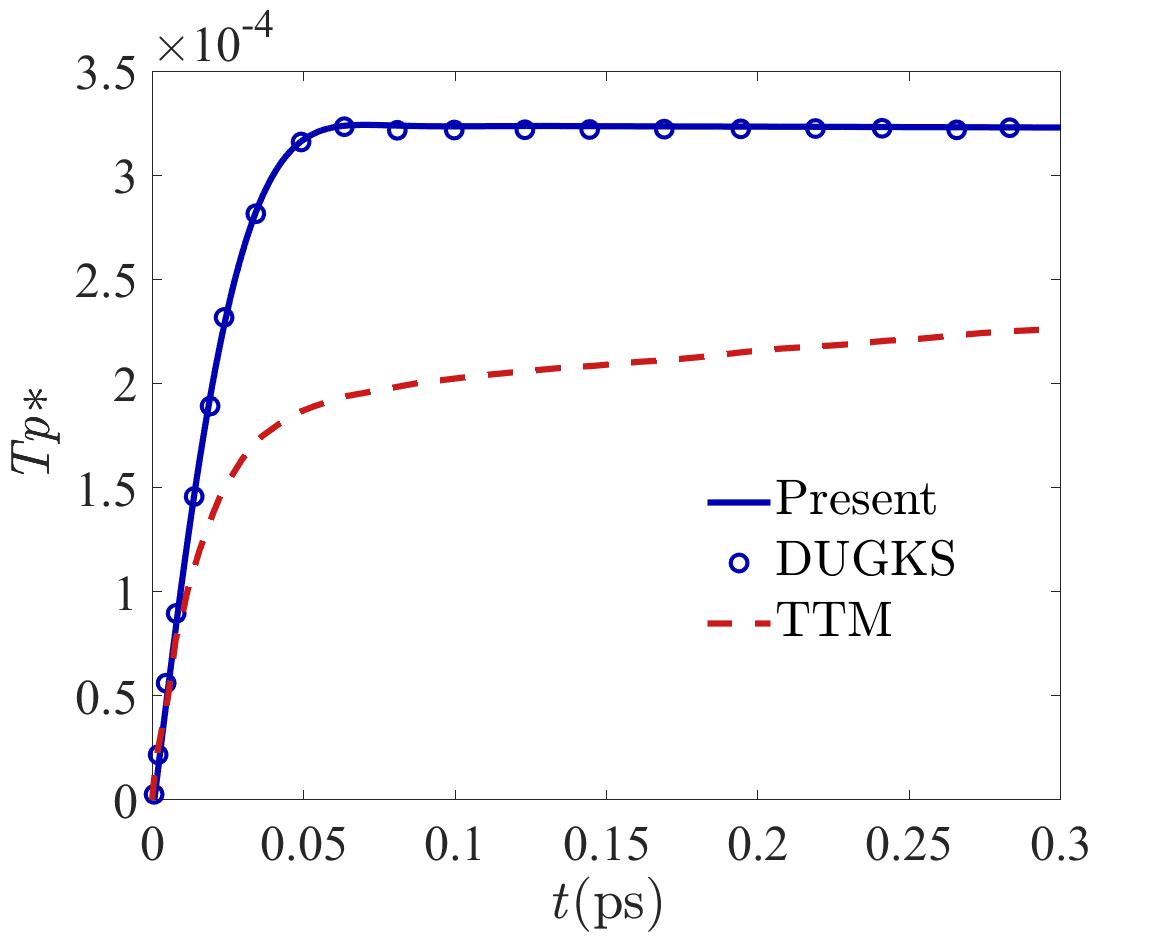}}~~ 
    \subfloat[$L$=1 $\mu$m, $\Delta x$=25 nm]{\label{fig3d}\includegraphics[width=0.40\textwidth]{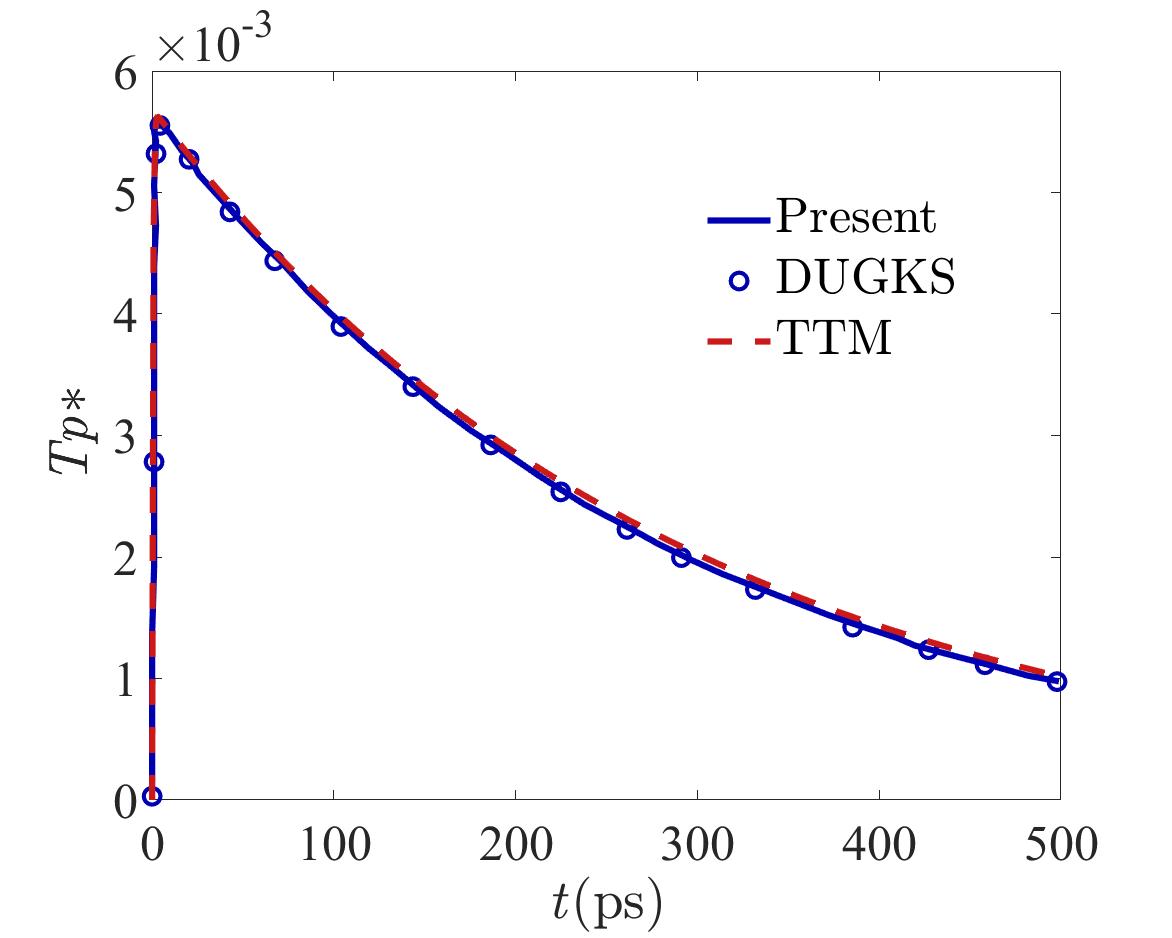}}\\ 
    \caption{Time-dependent electron and phonon temperatures at $x= L/4$ with different $L$, where $T_{ref}$= 300 K, $\Delta T$= 1 K, $T_{e}^*=\left ( T_e-{T}_{ref}\right )/\Delta T$, $T_{p}^*=\left (T_p-{T}_{ref}\right )/\Delta T$. The data predicted by the DUGKS and TTM is obtained from previous reference~\cite{ZHANG2024123379}. } 
    \label{fig3}
\end{figure}
\begin{figure}[h]
    \centering
    \subfloat[$L$=100 nm, $\Delta x$=2.5 nm]{\label{fig4a}\includegraphics[width=0.33\textwidth]{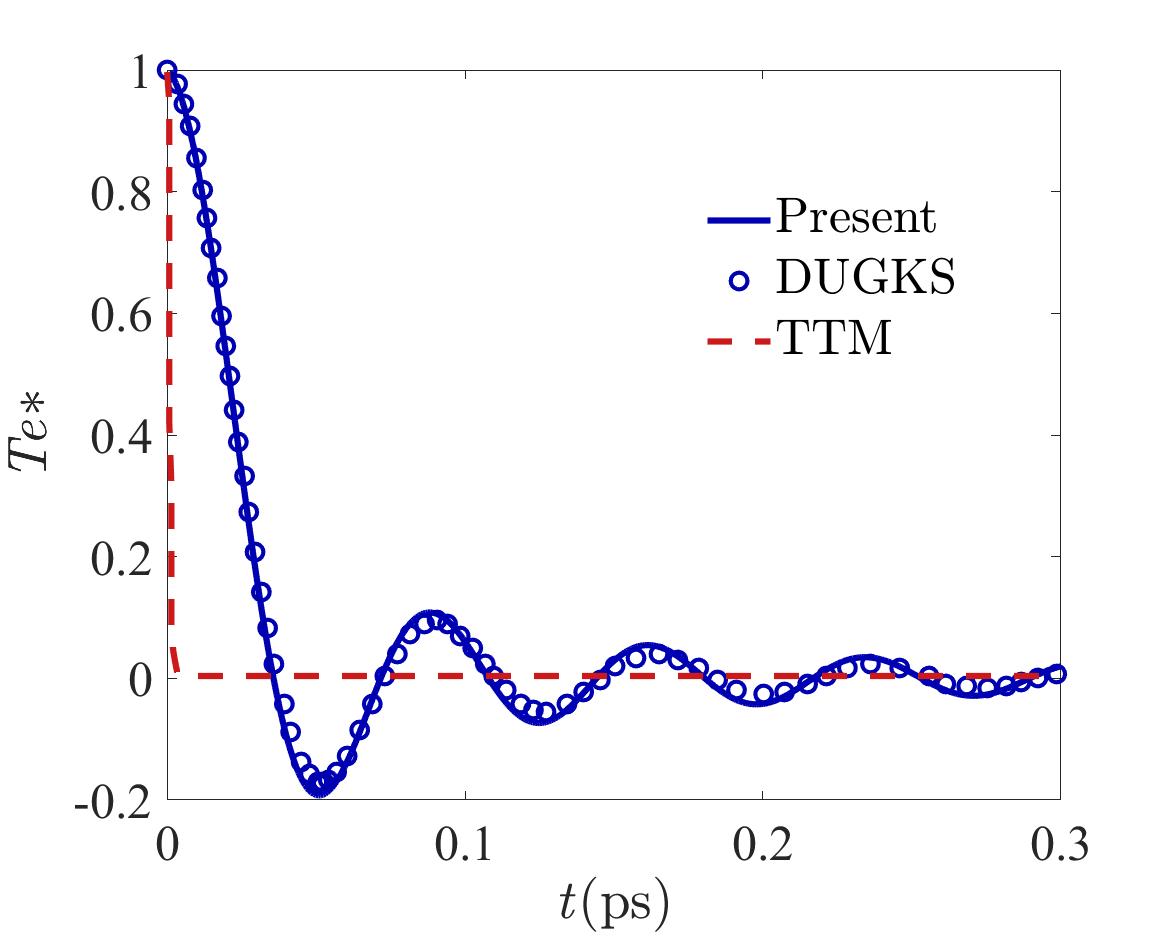}}~
    \subfloat[$L$=1 $\mu$m, $\Delta x$=25 nm]{\label{fig4b}\includegraphics[width=0.33\textwidth]{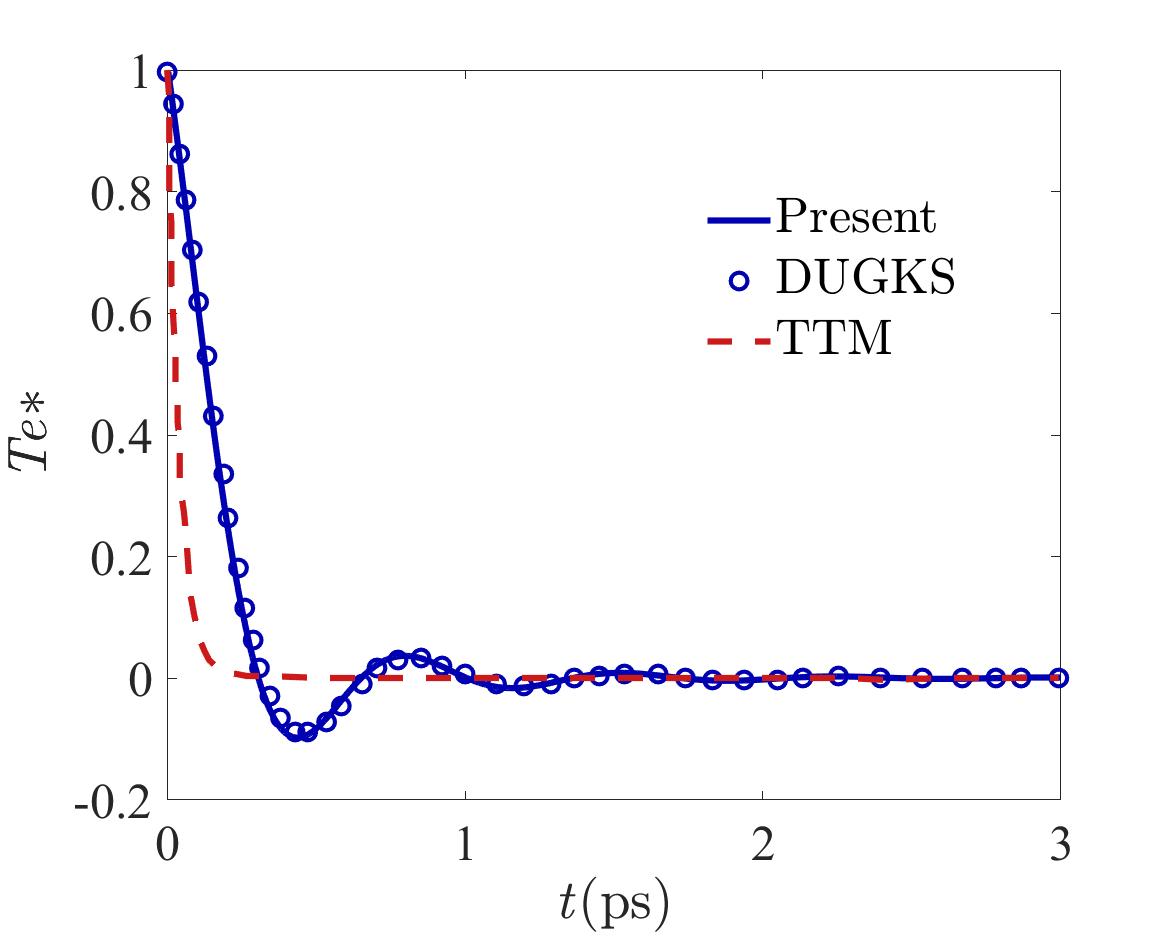}}~
    \subfloat[$L$=10 $\mu$m, $\Delta x$=250 nm]{\label{fig4c}\includegraphics[width=0.33\textwidth]{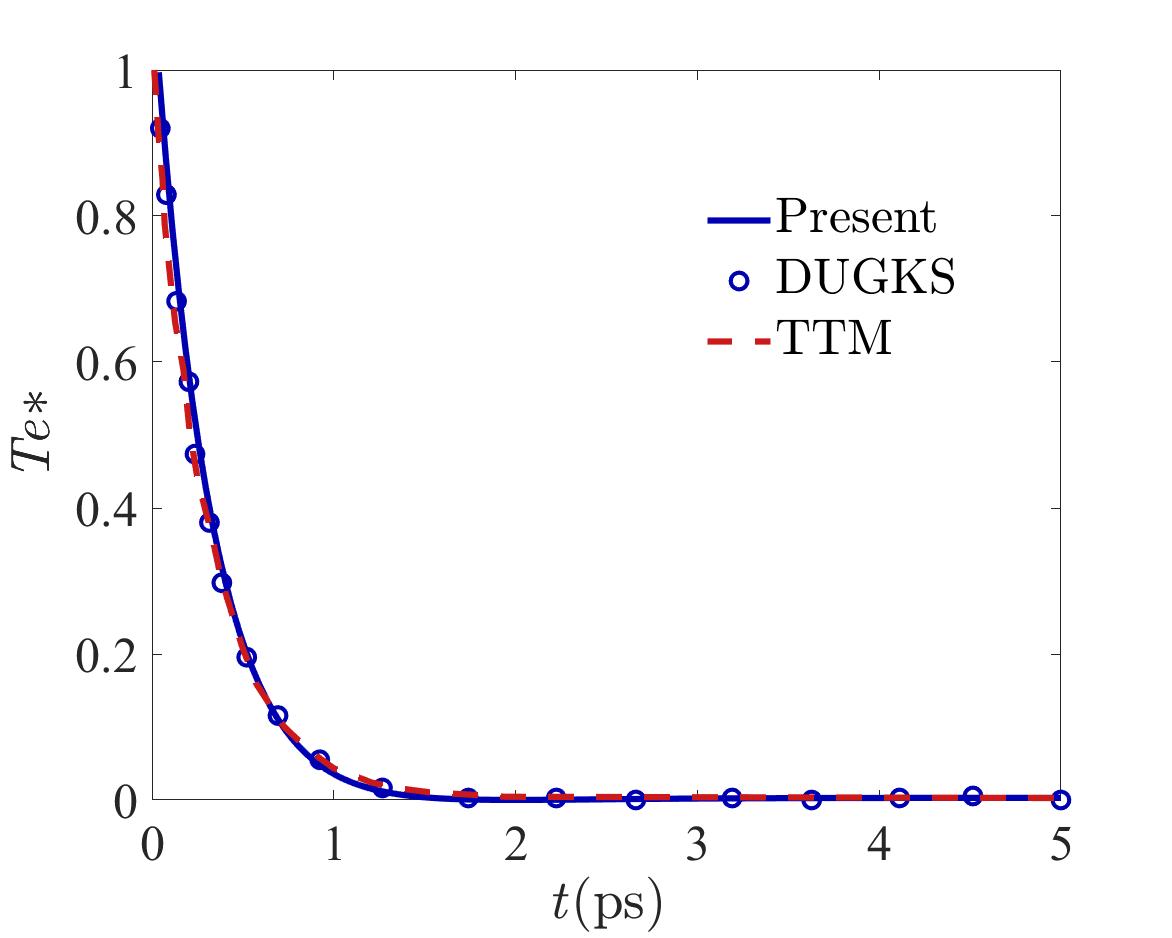}}\\ 
    \subfloat[$L$=100 nm, $\Delta x$=2.5 nm]{\label{fig4d}\includegraphics[width=0.33\textwidth]{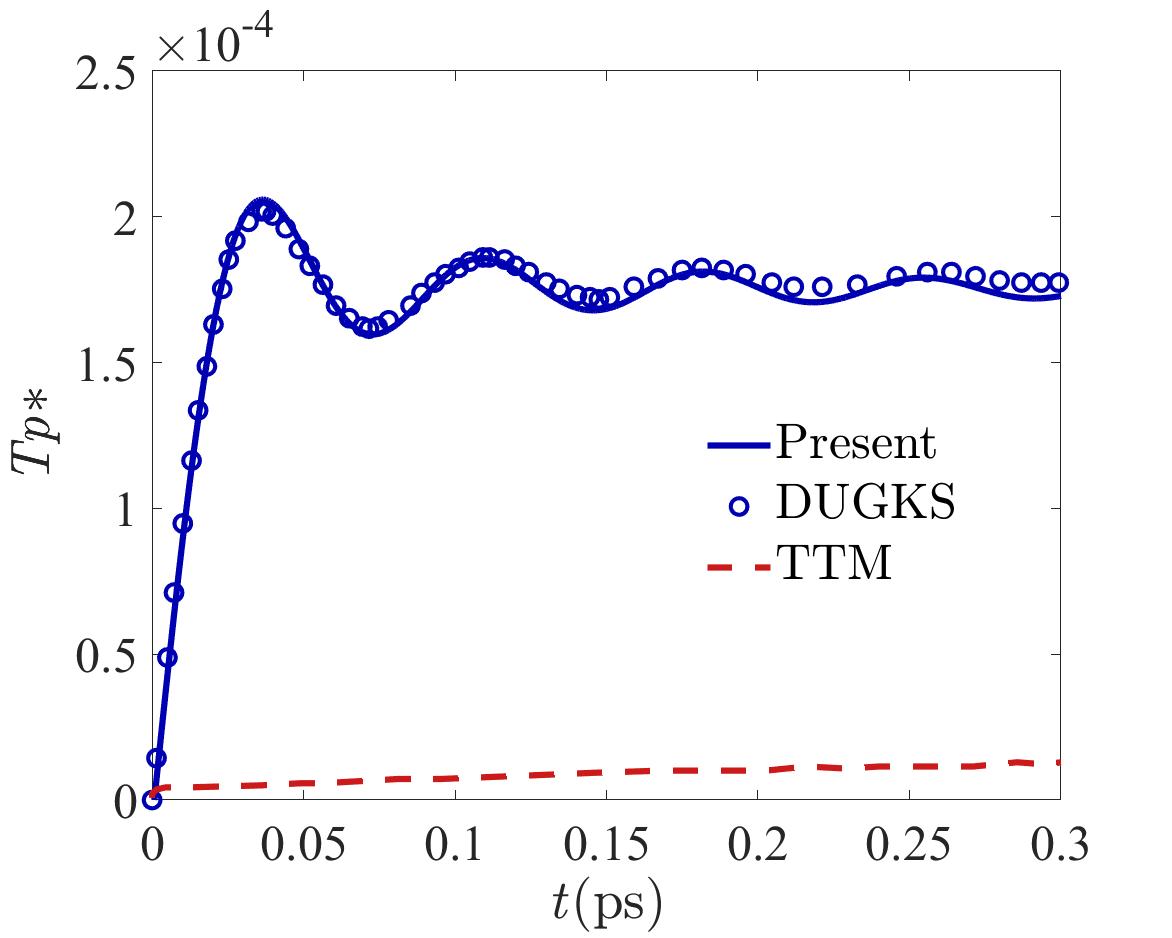}}~
    \subfloat[$L$=1 $\mu$m, $\Delta x$=25 nm]{\label{fig4e}\includegraphics[width=0.33\textwidth]{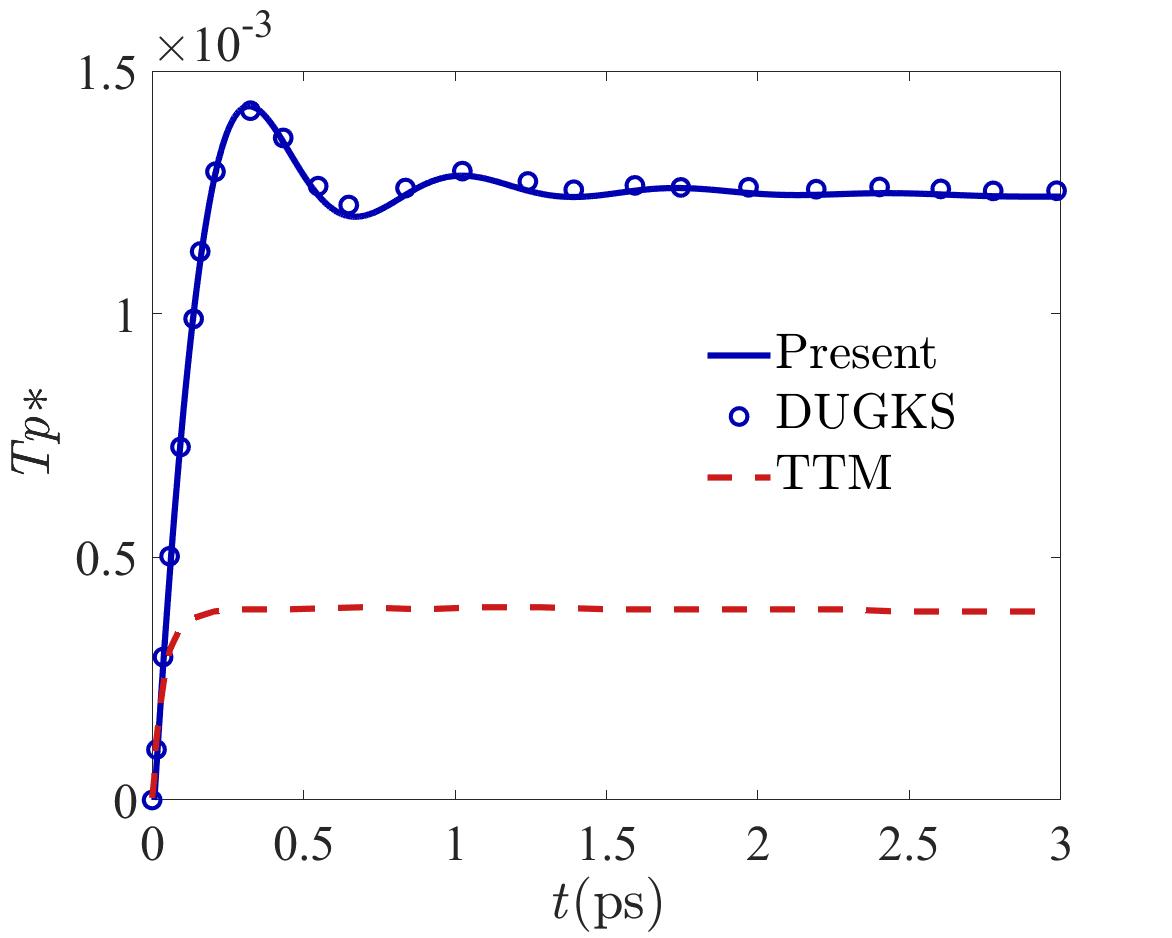}}~ 
    \subfloat[$L$=10 $\mu$m, $\Delta x$=250 nm]{\label{fig4f}\includegraphics[width=0.33\textwidth]{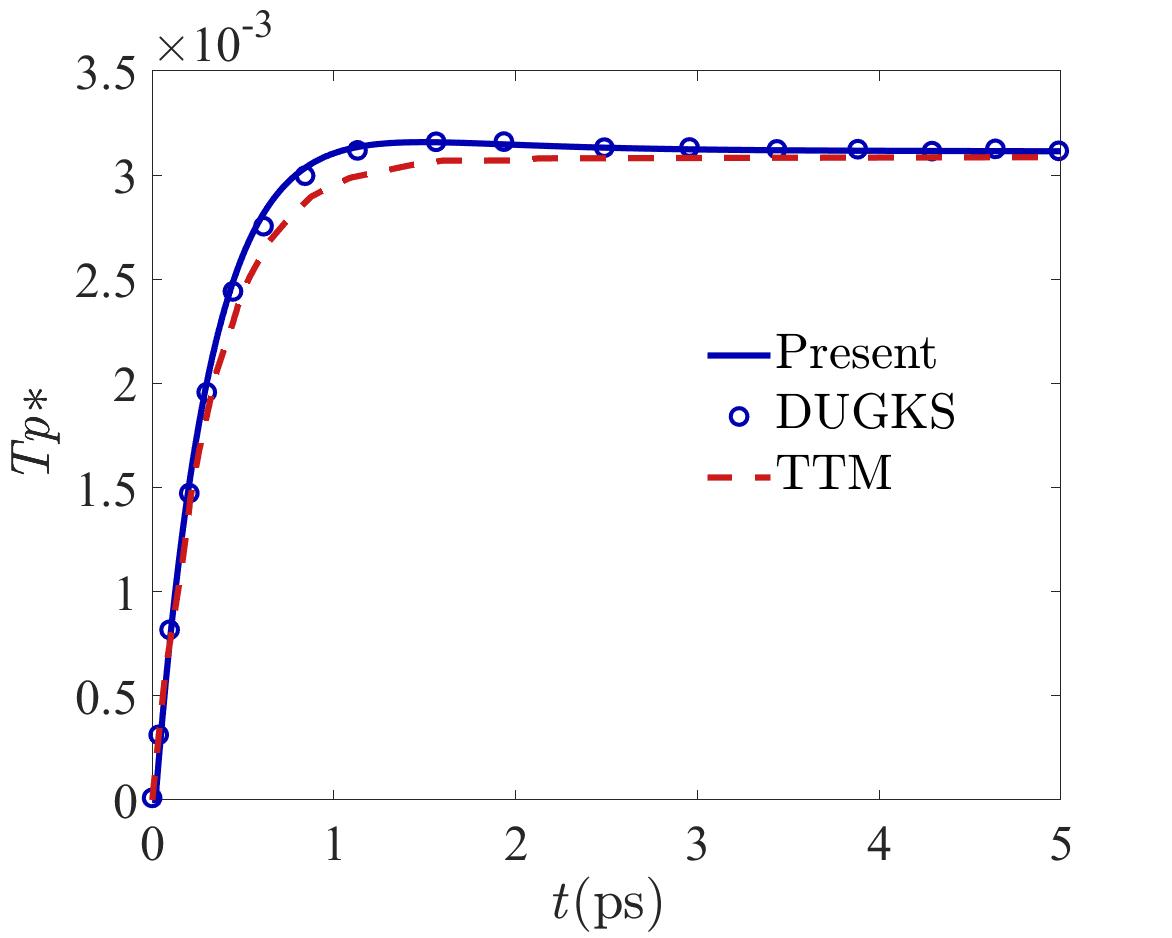}}
    \caption{Time-dependent electron and phonon temperatures at $x$= L/4 with different $L$, where $T_{ref}$= 25 K, $\Delta T$= 1 K, $T_{e}^*=\left ( T_e-{T}_{ref}\right )/\Delta T$, $T_{p}^*=\left ( T_p-{T}_{ref}\right )/\Delta T$. The data predicted by the DUGKS and TTM is obtained from previous reference~\cite{ZHANG2024123379}. }
    \label{fig4}
\end{figure}

The evolution of electron/phonon temperatures from nanometers to microns is systematically illustrated in~\cref{fig3,fig4} at normal or extremely low temperatures. 
It can be found that validation of the algorithm's accuracy is achieved through the close agreement of its results with those of the DUGKS across various scales. 
As demonstrated in~\cref{fig3a,fig3c,fig4a,fig4d} for the 100 nm scale, the predicted electron cooling rates, maximum phonon temperature increases, and their onset times are found to be nearly identical between the two methods.  
The curves are shown to be in substantial agreement at other larger scales.
Furthermore, it can be seen that at lower temperatures, the increase in the hot carrier mean free path leads to the time required for electron-phonon thermal equilibrium to be shortened, thereby enhancing the non-equilibrium ballistic effect . 
It is indicated that the strength of the non-equilibrium transport phenomenon is significantly influenced by temperature.

Significant deviations of the traditional TTM from present or DUGKS' results are exhibited at the nanoscale, while convergence toward agreement is observed at larger scales. 
For $L$=100 nm (\cref{fig3a,fig3c,fig4a,fig4d}), both the electron cooling rate and the peak phonon temperature are underestimated by TTM; 
as the scale is increased to 1 $\mu$m (\cref{fig3b,fig3d,fig4b,fig4e}), the deviation is reduced, though systematically low predictions are still maintained by TTM; at the 10 $\mu$m (~\cref{fig4c,fig4f}), consistency is largely achieved among all three methods. This indicates that TTM is applicable in regions where macroscopic diffusion is dominant, yet its limitations are revealed in ultra-fast non-equilibrium processes at the nanoscale.

The energy transfer process is visualized through a comparison of the electron and phonon temperature curves. A significantly shorter characteristic time is exhibited by electron temperature changes (cooling) compared to phonon temperature changes (heating). At small scales, the process is dominated by extremely rapid electron-phonon coupling, characterized by a plunge in electron temperature within sub-picoseconds and a sharp response in phonon temperature (\cref{fig3a,fig3c,fig4a,fig4d}). At the mesoscale, both temperatures are observed to evolve slowly over tens of picoseconds, showing distinct coupled relaxation characteristics (\cref{fig3b,fig3d,fig4b,fig4e}). At the macroscale, temperature changes occur very slowly, approaching classical diffusion behavior (\cref{fig4c,fig4f}). The complete relaxation chain following TTG excitation is revealed: energy is initially deposited in the electron subsystem, then transferred to the phonon subsystem via electron-phonon coupling, until thermal equilibrium is finally reached. 

In summary,the precise capture of heat transport behavior across multiple scales is achieved by the present algorithm. A high level of agreement with the results from the DUGKS is shown, while superior performance over the traditional TTM in modeling nanoscale ultra-fast non-equilibrium thermal processes is exhibited. 

\section{Conclusion}
\label{sec:conclusion}

A semi-implicit Lax–Wendroff kinetic scheme is developed, whereby a unified computational framework is provided for the thermal simulation of coupled electron–phonon transport across ballistic and diffusive regimes. 
The distinctive feature of this method is that it integrates the physical evolution information of heat carriers into the numerical modeling process. 
Specifically, the finite difference method is employed to resolve the kinetic model of electron-phonon coupling again when reconstructing the interfacial distribution function.
Consequently, the particle migration, scattering and electron–phonon coupling processes are coupled within a single time step, which successfully breaks the temporal and spatial constraints imposed by relaxation time and mean free path without the loss of solution accuracy.
The robustness and computational efficiency of this methodology for multiscale thermal conduction are validated through numerical tests. 
It is anticipated that this work will contribute to a more profound understanding of electron–phonon interactions and provide a refined instrument for thermal management within next-generation microelectronic devices.

\section*{Acknowledgments}

C. Z. acknowledges the support of the National Natural Science Foundation of China (52506078) and Zhejiang Provincial Natural Science Foundation of China under Grant No.LMS26E060012.
C. Z. acknowledges the members of online WeChat Group: Device Simulation Happy Exchange Group, and acknowledges Beijing PARATERA Tech CO., Ltd. for the HPC resources.
H. L. acknowledges the support of the National Natural Science Foundation of China (12572285).

\bibliographystyle{elsarticle-num-names_clear}
\bibliography{phonon}
\end{document}